\documentclass[times,twocolumn,final]{elsarticle}
\usepackage{cnf}
\usepackage{framed,multirow}

\usepackage{amssymb}
\usepackage{latexsym}
\usepackage{amsmath}
\usepackage[hyphens]{url}
\usepackage[hidelinks]{hyperref}
\usepackage{breakurl} % if using pdfLaTeX
\usepackage{xcolor}
\definecolor{newcolor}{rgb}{.8,.349,.1}

\usepackage{hyperref}

\usepackage[switch,pagewise]{lineno} %Required by command \linenumbers below

\journal{Combustion and Flame}

\begin{document}

\verso{H. Heng et al.}

\begin{frontmatter}

\title{Combustion Behaviour of Single Silicon Particles in Different Oxidizing Environments}%
\author[1,2]{Herman Heng\corref{cor1}}
\cortext[cor1]{Corresponding author: cloud.heng@mail.mcgill.ca (Cloud H. Heng)}
% \emailauthor{cloud.heng@mail.mcgill.ca}{Cloud Heng}
%\ead{example@email.com}
    
\author[2]{Hugo Keck}
\author[2]{Christian Chauveau}
\author[1]{Samuel Goroshin}
\author[1]{Jeffrey Bergthorson}
\author[2]{Fabien Halter}

\address[1]{McGill University, 817 Sherbrooke Street West, Montreal, Quebec H3A 0C3, Canada.}
\address[2]{ICARE-CNRS, 1C Avenue de la Recherche Scientifique, Orléans, 45100, France.}

\begin{abstract}
%%%
Silicon, despite its abundance and high energy density, remains underexplored as a carbon-free fuel, with limited data available on its combustion characteristics. In this work, the combustion behaviour of silicon particles is examined using an electrostatic levitator with laser ignition. Five oxidizing environments at atmospheric pressure are investigated: air, pure oxygen, and mixtures containing 40\% oxygen (by mole) diluted with nitrogen, helium, or argon. The burning droplet peak temperature, measured by three-colour pyrometry, increases by 337\,K from air to pure oxygen. The measured peak temperature of the silicon droplet in the 40\%O$_2$–60\%He mixture is noticeably lower than that in the 40\%O$_2$–60\%Ar mixture, in contradiction with thermodynamic predictions, due to a higher Lewis number of the helium-diluted mixture. Although oxygen diffusivity is higher in the helium-diluted mixture, a lower burning rate is observed, attributed to the lower combustion temperature. High-speed colour camera observations (with a resolution of about 1.25~\textmu m/px) reveal that the square of the particle diameter, $d^2$, decreases with time, $t$, in each individual combustion run, following a strong linear relationship ($R^2 > 0.99$) across all oxidizing environments. However, the combustion lifetime, $\tau_\mathrm{c}$, and initial particle diameter, $d_\mathrm{i}$, follow a trend of $\tau_\mathrm{c} \propto d_\mathrm{i}^{n}$, with $n$ ranging from 1.69 to 1.82. This deviation from the expected $n = 2$ appears to result from unavoidable measurement uncertainties and the limited particle size range, rather than differences in combustion physics. The decrease in silicon droplet size during combustion is attributed to the formation of gaseous SiO as an intermediate combustion product. The SiO species is observed using a UV camera, showing a UV intensity decay from the particle surface. High-speed imaging and LED absorption signals indicate that the final condensed product, SiO$_2$ nanoparticles, are not optically visible, suggesting they possess very low emissivity.

% The burning rate constant is found to deviate from the theoretical prediction likely due to the vaporization of Si at high combustion temperature. 
% The SiO boundary layer thickness shows no significant difference between air and pure oxygen environments and decreases only gradually over time. 
%%%%
\end{abstract}


\end{frontmatter}

%\linenumbers

\section*{Novelty and significance statement}

We present the first experimental demonstration of ignition and complete combustion of single silicon particles, providing the first evidence that their combustion follows the classical $d^2$ law, unlike many previous studies on metal particles that reported deviations. The identified influence of Lewis number and particle size on combustion time, as well as the observed radiative properties of silicon combustion products, offers insights of scientific and engineering significance for developing combustion technologies using silicon as a carbon-free energy carrier.

%% main text
\section{Introduction \label{sec:intro}}
Metal fuels have gained increasing attention over the past few decades as carbon-free energy carriers. Silicon is a promising candidate in this role due to its abundance in Earth's crust and high specific energy density. Its widespread availability makes it attractive, potentially eliminating the need to recycle its combustion product (silicon dioxide) for reduction. Instead, the SiO$_2$ nano-particle product generated during combustion can be repurposed for material synthesis in industries such as cement, ceramics, and reflective coatings \cite{vaiani_ceramic_2023, american_coatings_association_use_2020} -- integrating clean energy production with local manufacturing. These key attributes may position silicon as a potentially more favourable candidate than aluminum and iron, the most extensively studied metal fuels in the literature to date.

Despite these advantages, silicon has not yet been widely considered as fuel within the combustion community. Most prior applications of silicon combustion have focused on pyrotechnics \cite{ai-kazraji_fast_1979, rugunanan_reactions_1991, koch_special_2007, conkling_chemistry_2019} rather than investigating its potential and properties in the context of sustainable fuels for energy systems. While silicon has been investigated as a component in composite propellants \cite{zuo_thermal_2022}, combustion data on pure silicon metal remains limited. This is due to the challenge of igniting silicon particles, caused by a protective silicon dioxide layer \cite{weiser_ignition_2003}, and its relatively slow combustion rate compared to more volatile metals, such as aluminum and magnesium. Nevertheless, recent work at McGill University \cite{heng_silicon_2025} has successfully demonstrated the first laminar silicon dust flame stabilized on a Bunsen burner. This breakthrough not only confirms that pure silicon can sustain self-propagating combustion for energy applications but also enables detailed investigation into its combustion characteristics, including laminar burning velocity, flame temperature, combustion products, and the combustion mode of silicon. Nevertheless, dust flame experiments still present limitations, particularly in observing the behaviour of individual particles, such as their size and temperature evolution during combustion. These insights are essential for deepening our understanding of the combustion process and for validating the applicability of existing available theoretical combustion models to silicon particle combustion.

The classical $d^2$ law of combustion \cite{turns_introduction_2000} has long been used to describe the temporal evolution of a burning droplet’s size for diffusion-controlled combustion, as expressed in Equation~\ref{eq:d2law}:
\begin{equation}
d^2 = d_\mathrm{i}^2 - Kt
\label{eq:d2law}
\end{equation}
where $d$ is the instantaneous particle diameter, $d_\mathrm{i}$ is the initial diameter, $t$ is time, and $K$ is a proportionality constant. When the particle is fully consumed ($d$ = 0), Equation~\ref{eq:d2law} reduces to $\tau_\mathrm{c} = d_\mathrm{i}^2 / K$, where $\tau_\mathrm{c}$ is the complete combustion time. This relationship implies that the combustion time is directly proportional to the square of the initial particle diameter, with the proportionality constant $K$ characterizing the burning rate. The $d^2$ law has been observed in experiments for conventional fuels, such as liquid hydrocarbons \cite{nuruzzaman_combustion_1971, faeth_fuel_1971} and char \cite{asheruddin_analysis_2022}, for most of the droplet particle combustion lifetime. In the case of metal fuels, some studies have shown that aluminum combustion somewhat fits the trend of the $d^2$ law \cite{marion_studies_1996, dreizin_mechanism_1999, braconnier_detailed_2019}; however, it is also recognized that the formation of an aluminum oxide cap on the droplet surface significantly alters the symmetry of aluminum droplet combustion, which affects the accuracy of the classical model. Several models have been proposed to account for the effects of aluminum oxide cap formation \cite{liu_modeling_2021, wang_modeling_2021, chen_exploring_2025} in the estimation of burning time. Iron particle combustion presents a different heterogeneous combustion mode. Because the boiling points of iron and its (sub-)oxides are typically lower than the flame temperature, the particle size does not necessarily decrease as predicted by the classical $d^2$ law. Ning et al. \cite{ning_burn_2021} reported a $d^n$-law with $n = 1.46-1.72$ for single iron particles, indicating a deviation from the classical $d^2$ law. Several studies have attempted to develop a new model to account for particle growth and changes in particle density after oxidation \cite{ning_experimental_2024, ning_quantitative_2024} in iron particle combustion.

There have been numerous numerical and experimental investigations on the combustion of single iron and aluminum particles within the combustion community. However, for silicon, combustion data remain scarce. As a result, the combustion behaviour of individual silicon particles is still not well understood. It should be noted that metal fuels are more chemically diverse than hydrocarbon fuels \cite{goroshin_fundamental_2022}, where insights gained from the combustion of one metal fuel cannot be directly applied to another. Each metal fuel requires its own comprehensive investigation and understanding, as their reaction chemistry, thermodynamics, and phase transformations are fundamentally distinct without a unified framework. 

In the present study, using an electrostatic levitator coupled with several diagnostic techniques, we successfully stabilize and ignite spherical silicon particles. We present the first comprehensive experimental dataset for single-particle silicon combustion, including time-resolved measurements of particle temperature in different oxidizing gas mixtures. Unlike our previous silicon dust flame experiments \cite{heng_silicon_2025} performed at 30–40\% oxygen concentration, the single-particle setup allows operation with up to 100\% oxygen. Leveraging this advantage, we perform a broader investigation of how combustion temperature varies with oxygen concentration and with different inert diluting gases. We also aim to examine the evolution of the particle diameter in real time during combustion in various oxidizing environments. We further analyze the combustion time to assess whether silicon combustion follows the classical $d^2$ law of combustion. Finally, we identify the presence of a SiO boundary layer surrounding the burning droplet and investigate the behaviour of the final condensed combustion products through analysis of the light absorption signal.

% across five oxidizing environments with varying oxygen concentrations and diluent gases.
% and discuss whether modifications to the classical model are necessary
% From this relationship, we determine the burning rate constant of silicon particles and compare it with theoretical estimates.

\section{Experimental Setup \label{sec:setup}}

\subsection{Silicon Powder and Gases} \label{ss:powder_and_gas}
\begin{figure}[h]
\centering
\includegraphics[scale=0.3]{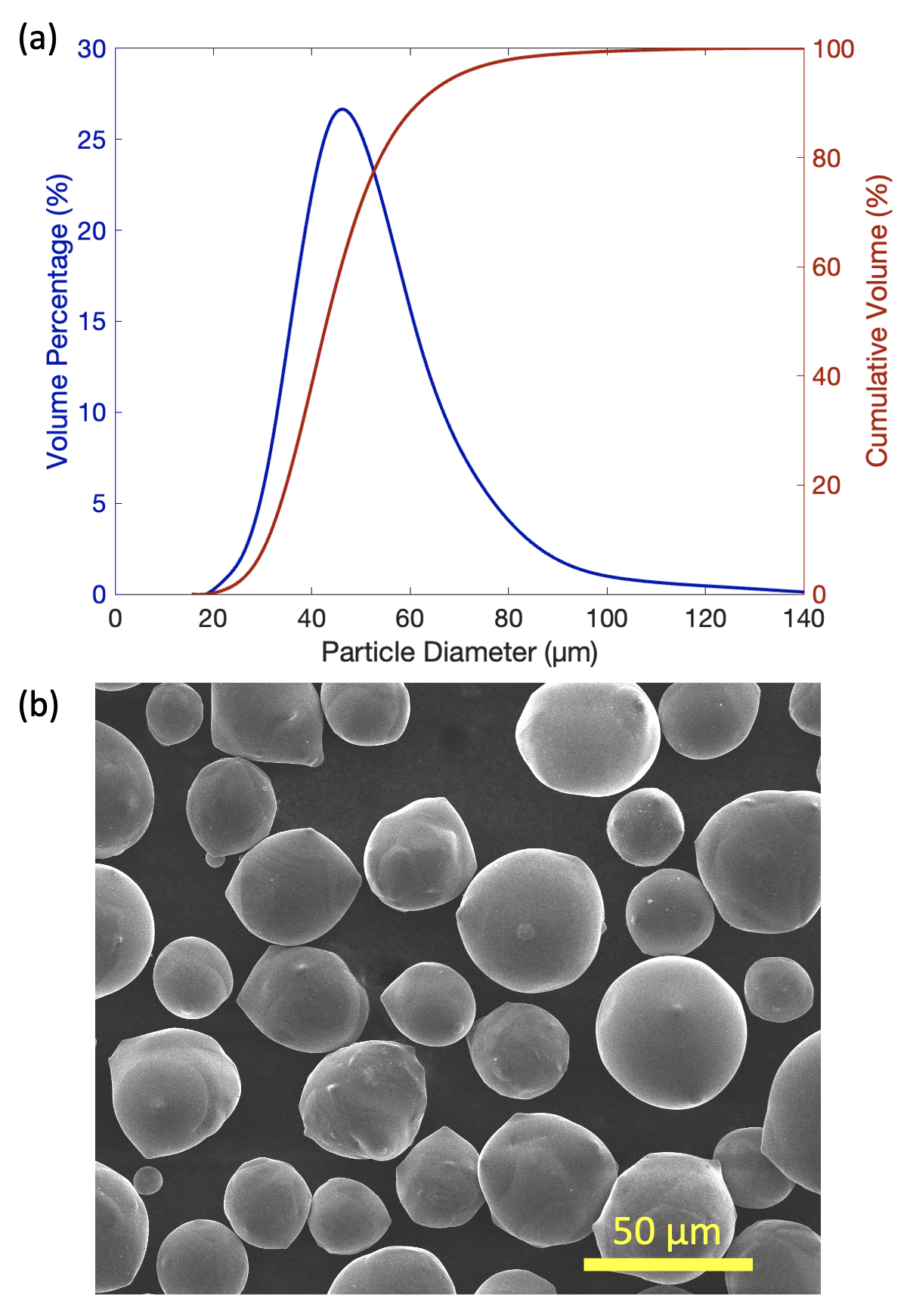}
\caption{(a) Particle size distribution of the silicon powder (b) A sample image obtained through scanning electron microscopy (SEM) of the silicon powder \label{fig:PSDnSEM}}
\end{figure}

The silicon powder is supplied by Nanochemazone. The powder is spherical in shape, with a mean Sauter diameter of 45.6~\textmu m. The particle size distribution, measured by laser diffraction, and particle morphology, captured by scanning electron microscopy (SEM), are shown in Fig.~\ref{fig:PSDnSEM}. In this study, silicon particle combustion is investigated in five oxidizing environments: air, pure O\textsubscript{2}, and a fixed oxygen concentration of 40\% (by mole) diluted with nitrogen, helium, and argon, respectively. The ambient pressure was maintained at 1\,atm for all experimental runs.

\subsection{Electrostatic Levitator}

The present single silicon particle combustion experiments are performed using an electrostatic levitation facility at the Institut de Combustion, Aérothermique, Réactivité et Environnement (ICARE-CNRS), Orléans, France. The concept of electrodynamic levitation for the study of particles and droplets was developed in the 1980s of the last century~\cite{bar-ziv_electrodynamic_1991}, and the ICARE-CNRS levitation facility has since been used for combustion studies of several metals and metal alloys. For further details of the setup and earlier experimental work, readers are referred to Legrand et al.~\cite{legrand_ignition_1998} for magnesium and magnesium–aluminum alloys in CO$_2$, and to Marion et al.~\cite{marion_studies_1996} and Braconnier et al.~\cite{braconnier_detailed_2019} for aluminum burning in air at varying ambient pressure conditions. 

In brief, the electrostatic levitator generates some electric fields through a combination of DC and AC voltage supplies on three electrodes. These electric fields counteract both gravity and radial forces acting on the particle, allowing for stable levitation of a single charged silicon particle at a fixed position without significant motion. Once stabilized, the particle is ignited using a focused 50-watt CO\textsubscript{2} laser beam. The laser beam position is carefully calibrated such that the stabilized silicon particle is located at the peak of the laser’s Gaussian intensity profile. The levitator can be operated in ambient air or enclosed in a chamber filled with the specific gas mixtures outlined in Section~\ref{ss:powder_and_gas}. 

\begin{figure}[h]
\centering
\includegraphics[scale=0.45]{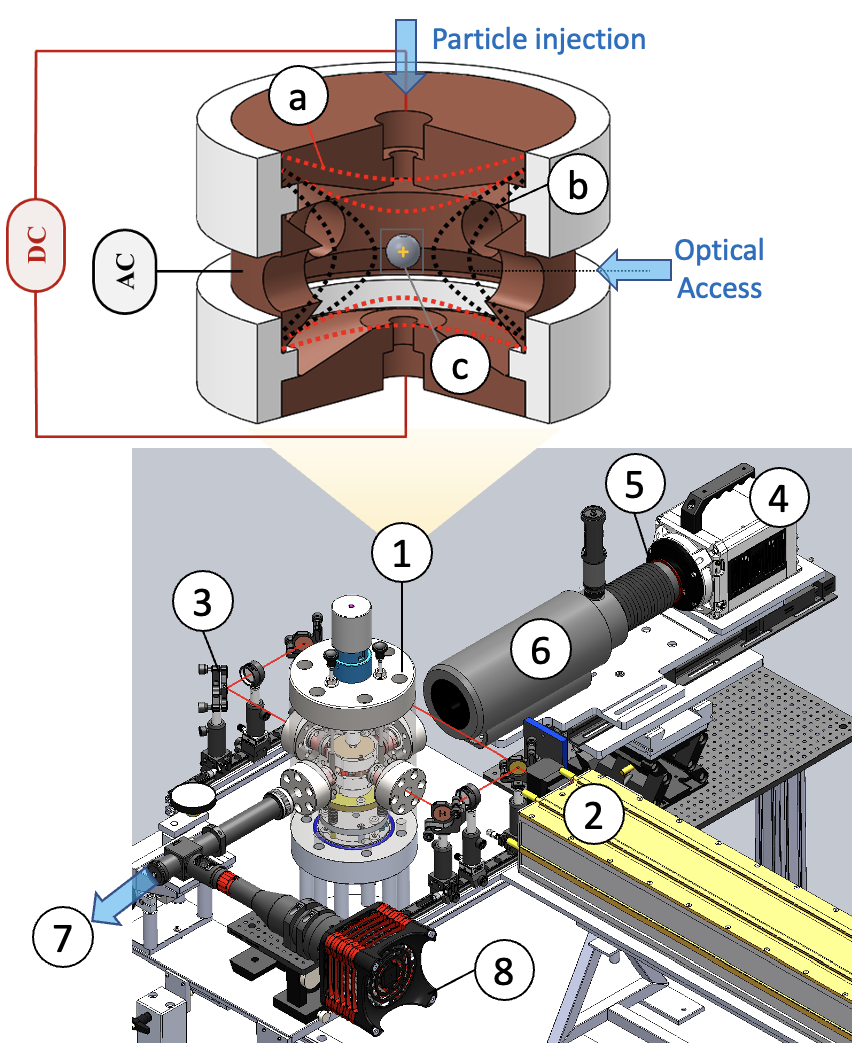}
\caption{(1) The electrostatic levitator with (a) gravity-compensating fields (red dotted lines) and (b) centering fields (black dotted lines) that stabilize (c) single charged silicon particles in the setup, (2) CO\textsubscript{2} laser system with a set of (3) optical mirrors used to reflect the laser beam for a more uniform laser distribution on the particle, (4) colour or UV camera coupled with a (5) bandpass filter and a (6) telephoto lens, (7) light emission transmitted towards a photomultiplier system, (8) white LED used as a backlight. \label{fig:setup}}
\end{figure}

\subsection{Optical Diagnostics}
Figure \ref{fig:setup} shows the schematic of the electrostatic levitator and the optical diagnostic tools used in the experiment. During combustion, light emitted by the burning silicon particle is recorded using a Thorlabs PMT1002 photomultiplier system, with three wavelength channels set at 700\,nm, 760\,nm, and 820\,nm. This three-colour pyrometry system is calibrated using a blackbody source at 1773\,K, from which sensitivity ratios between each channel pair are determined under varying light exposures. During experiments, the amount of light transmitted to the photomultiplier is regulated via an aperture at the entry of component\,7 in Fig.~\ref{fig:setup}, ensuring that the voltage signal produced falls within the linear response range of the pyrometer and does not approach saturation.

The voltage signal from the photomultiplier is also used to automatically shut off the ignition laser once the detected voltage signal exceeds a predefined threshold, thereby minimizing laser interference with the combustion process. To capture combustion dynamics, a Phantom T2410 colour high-speed camera, equipped with a triple bandpass filter (457\,nm, 530\,nm, 628\,nm), is used to record real-time particle imagery. For experiments in air, a Phantom T3610 UV camera is also used, equipped with a bandpass filter in the 250–300\,nm range. These cameras are coupled with a telephoto lens, which yield a resolution of about 1.25 \textmu m/pixel. The high-speed camera and photomultiplier system are synchronized to allow direct comparison of emission signals, derived temperatures, and visual recordings. A Thorlabs SOLIC-3C white LED system, developed from the recent work of Keck et al.~\cite{keck_new_2024, keck_temperature_2025}, is used as a backlight to visualize the particle prior to ignition and to enable observation of absorption signals during combustion. The LED emits light in the 400–750\,nm wavelength range.

\section{Results and Discussion}

\subsection{Temperature evolution}\label{subsec:temp}

\begin{figure*}[t]
\centering
\includegraphics[scale=0.48]{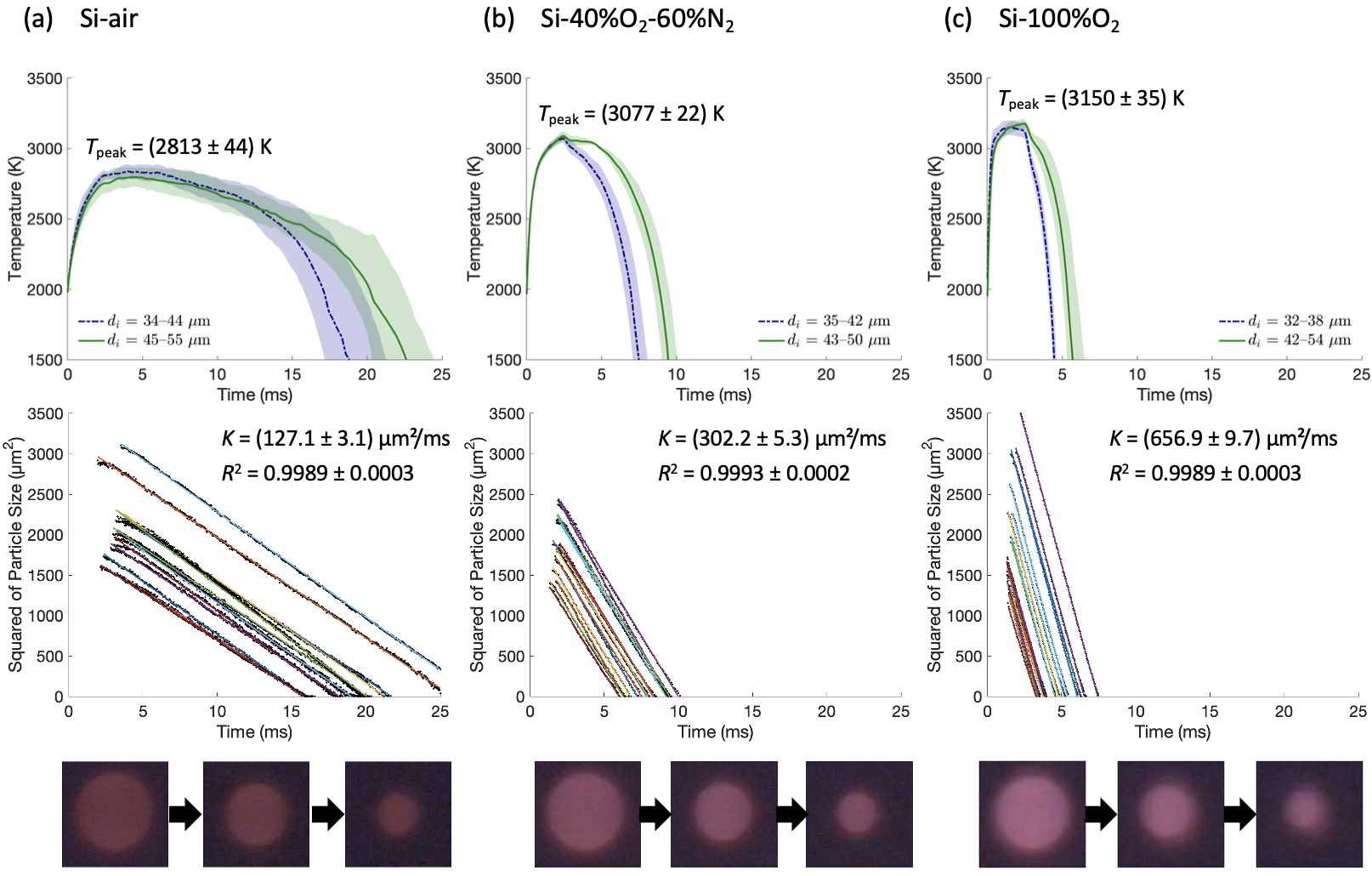}
\includegraphics[scale=0.48]{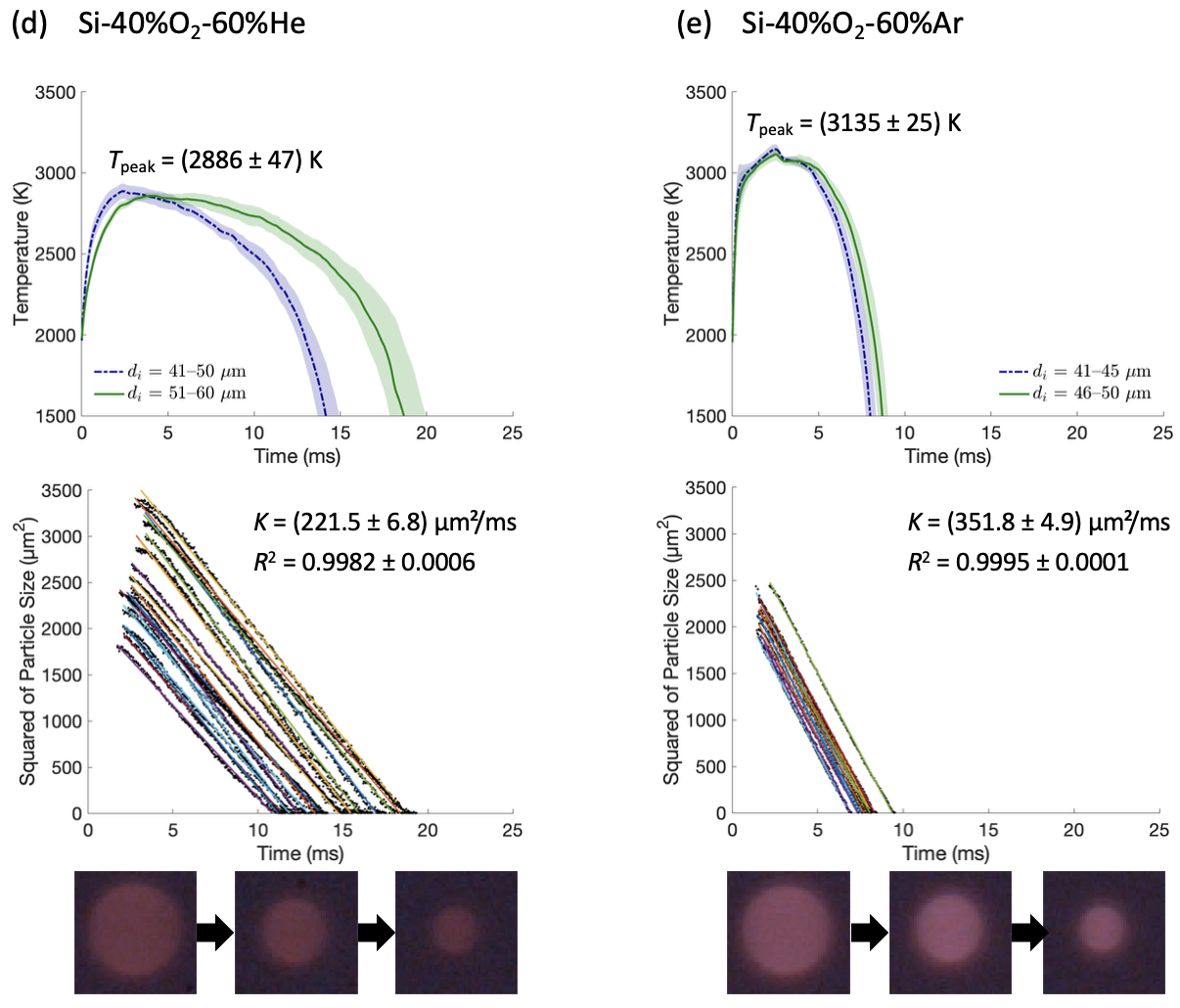}
\caption{Temperature and particle size evolutions in different oxidizing environments: (a) air, (b) 40\%O$_2$–60\%N$_2$, (c) 100\% O$_2$, (d) 40\%O$_2$–60\%He, and (e) 40\%O$_2$–60\%Ar. For each sub-figure, the top row shows the temperature evolution over time with respective standard errors of means for two initial silicon particle size ranges, while the middle row illustrates the evolution of the square of particle diameter over time for each experimental run with the slope and $R^2$ values shown for the fitted lines. The bottom row provides three sample frames of the burning silicon droplet, and the sample frames shown in all the sub-figures are taken at the same camera exposure settings.
\label{fig:tempNsize}}
\end{figure*}

The voltage signals measured by the three-colour pyrometer are corrected using the sensitivity ratios obtained from calibration. Based on the corrected relative signals for each wavelength channel pair, the temperature of the condensed phase (i.e., the burning silicon droplet) can be derived using Planck’s law assuming that it is a grey body. The resulting temperature profiles are shown in the top row of Fig.~\ref{fig:tempNsize}(a)-(c) for three oxygen concentrations: air, 40\% O\textsubscript{2} diluted in nitrogen, and pure O\textsubscript{2}. The top row of Fig.~\ref{fig:tempNsize}(d)-(e) shows the temperature profiles for 40\% O\textsubscript{2} diluted in helium and argon, respectively. 

% For each gas mixture, the temperature sharply rises following the laser ignition and reaches a peak. 

Comparing the peak temperatures across different oxygen concentrations, the experimental results show that, for a wide range of oxygen concentrations from air to pure O\textsubscript{2}, the measured peak temperature increases by 337\,K. This temperature rise aligns with the thermodynamic equilibrium analysis using CEARUN \cite{gordon_computer_1996}, which predicts an adiabatic flame temperature increase of approximately 282\,K when the oxygen concentration increases from 21\% to 100\% in nitrogen gas. Similar to conventional fuels, the combustion temperature is expected to increase due to the reduction of inert species that absorb the produced heat from combustion. However, the observed temperature rise of 337\,K for silicon from air to pure oxygen is relatively small compared to that of conventional fuels (e.g., methane or hydrogen). This lower sensitivity is due to the fact that the silicon combustion temperature appears to favour the formation of SiO instead of SiO\textsubscript{2} as predicted by thermodynamic equilibrium calculations. The enthalpy of formation for the reaction $\mathrm{Si} + 0.5\mathrm{O_2} \rightarrow \mathrm{SiO}$ is –100.4 kJ/mol, which is approximately nine times lower in magnitude compared to that of the reaction $\mathrm{Si} + \mathrm{O_2} \rightarrow \mathrm{SiO_2}$, which has an enthalpy of formation of –905.5 kJ/mol \cite{nist_nist_2025}. Since SiO is the favoured product under typical silicon combustion temperatures, it is expected that less heat energy is released and more chemical energy is retained in the reaction zone, resulting in a lower sensitivity of combustion temperature to oxygen concentration. This temperature insensitivity to oxygen concentration has also been observed in the previous experimental study on silicon dust flames \cite{heng_silicon_2025}. Despite the relatively modest temperature rise (337\,K), the luminosity ($I$) of the burning silicon droplet increases noticeably, as shown in the sample images of the burning droplet in Fig.~\ref{fig:tempNsize}(a)-(c), due to its strong dependence on temperature ($I \propto T^4$).

% \begin{figure*}[!t]
% \centering
% \includegraphics[width=1\textwidth]{figures/temp_O2.png}
% \caption{Top row: Temperature evolution over time with respective standard errors of means for two initial silicon particle size ranges. Bottom row: Evolution of the squared particle diameter over time for each experimental run with the slope and $R^2$ values shown for the fitted lines. Both evolutions are investigated in different gas mixtures with varying oxygen concentrations: (a) air, (b) 40\%O$_2$–60\%N$_2$, and (c) 100\% O$_2$ with three sample frames of the silicon droplet shown for each gas mixture. 
% \label{fig:tempNsize_O2}}
% \end{figure*}

The peak temperatures of the silicon droplet in different diluting gases — helium, nitrogen, and argon — are compared in Fig.~\ref{fig:tempNsize}(b), (d), and (e), respectively, with the oxygen concentration fixed at 40\% by mole. According to thermodynamic equilibrium calculations, given the same fuel, equivalence ratio, and oxygen concentration, the adiabatic temperatures for helium and argon as diluents are identical, since both inert gases are monoatomic with identical molar heat capacities. However, as shown in Fig.~\ref{fig:tempNsize}(d) and (e), the measured peak temperature in the gas mixture of 40\%O$_2$–60\%Ar is noticeably higher than that in the gas mixture of 40\%O$_2$–60\%He, which deviates from the thermodynamic analysis. 

% Furthermore, if the Lewis number is unity ($ \mathrm{Le} = \alpha / D_{\mathrm{O}_2\mathrm{-mix}} = 1 $), the measured combustion temperature in the case of heat loss should be also similar for both gas mixtures. This is because, in a diffusion-controlled combustion process, the ratio of heat generation rate (governed by the oxygen mass diffusivity in the mixture: $Q_\mathrm{G} \propto D_{\mathrm{O}_2\mathrm{-mix}}$) to heat loss rate (proportional to thermal diffusivity: $Q_\mathrm{L} \propto \alpha$) remains constant. 

% \begin{figure}[h]
% \centering
% \includegraphics[scale=0.4]{figures/temp_diluent.png}
% \caption{(a) Sample frames of a burning silicon droplet, (b) temperature evolution over time, (c) evolution of the squared particle diameter over time. Both evolutions are examined with a fixed oxygen concentration of 40\% O$_2$, while each gas mixture is diluted in He, N$_2$, and Ar, respectively. All initial particle sizes are in a narrow range of 41–50 ~\textmu m.
% \label{fig:tempNsize_diluent}}
% \end{figure}

For diffusion-limited combustion, the heat generation rate $\dot Q_\mathrm{G}$ is proportional to the rate of oxygen mass transport and the heat released at the particle surface, as expressed in Equation~\ref{eq:heat_gen}. The heat loss rate $\dot Q_\mathrm{L}$ is given by Equation~\ref{eq:heat_loss}, assuming that it occurs solely through conductive/convective transport. 

\begin{equation}
\dot Q_\mathrm{G} \propto \mathrm{Sh} \frac {D_{\mathrm{O_2}-\mathrm{mix}}}{r_\mathrm{p}}A_\mathrm{p}q \nu X_{\mathrm{O_2,\infty}}
\label{eq:heat_gen}
\end{equation}

\begin{equation}
\dot Q_\mathrm{L} = \mathrm{Nu} \frac{\lambda}{r_\mathrm{p}}A_\mathrm{p}(T_\mathrm{p} - T_\mathrm{g})
\label{eq:heat_loss}
\end{equation}

\noindent Here, $\mathrm{Sh}$ is the Sherwood number, $\mathrm{Nu}$ is the Nusselt number, $D_{\mathrm{O_2}-\mathrm{mix}}$ is the diffusivity of oxygen in the gas mixture, $\lambda$ is the thermal conductivity of the gas mixture, $r_\mathrm{p}$ is the particle radius, $A_\mathrm{p}$ is the surface area of particle, $q$ is the heat of reaction per mole of fuel, $\nu$ is the stoichiometric coefficient, $X_{\mathrm{O_2, \infty}}$ is the oxygen mole fraction in the bulk gas, $T_\mathrm{p}$ is the particle temperature, and $T_\mathrm{g}$ is the gas temperature. By applying an energy balance (i.e., equating heat generation and heat loss) and substituting $\lambda = \alpha \rho C_\mathrm{P}$, where $\alpha$, $\rho$, and $C_\mathrm{P}$ are the thermal diffusivity, density, and specific heat capacity at constant pressure of the gas mixture, respectively, and using the definition of the Lewis number, $\mathrm{Le} = \alpha / D_{\mathrm{O_2}\text{-}\mathrm{mix}}$, the following relationship is obtained:

\begin{equation}
T_\mathrm{p} - T_\mathrm{g} \propto \frac{1}{\mathrm{Le}} \cdot \frac{\mathrm{Sh}}{\mathrm{Nu}} \cdot \frac{q \nu X_{\mathrm{O_2, \infty}}}{\rho C_\mathrm{P}}
\label{eq:particle_temp}
\end{equation}

\begin{figure*}[h]
\centering
\includegraphics[width=1 \textwidth]{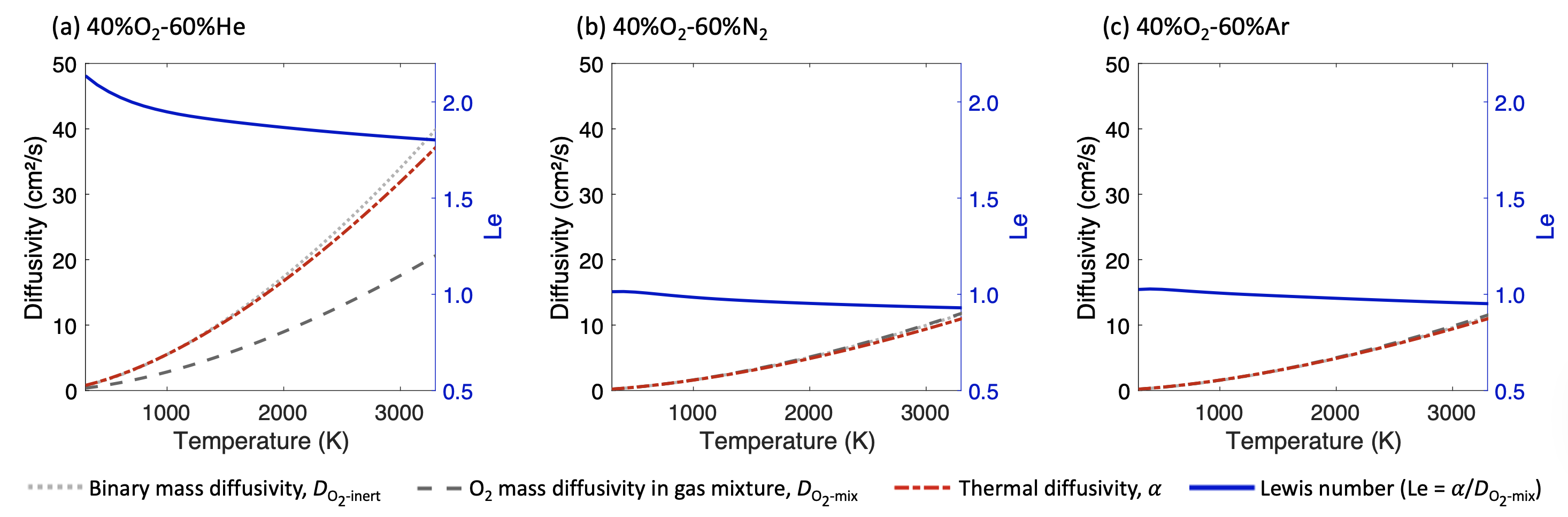}
\caption{Mass and thermal diffusivity at varying temperature and the corresponding Lewis number in three gas mixtures: (a) 40\%O$_2$–60\%He, (b) 40\%O$_2$–60\%N$_2$, and (c) 40\%O$_2$–60\%Ar.
\label{fig:leCANTERA}}
\end{figure*}

The term $q \nu X_{\mathrm{O_2, \infty}}/{\rho C_\mathrm{P}}$ in Equation~\ref{eq:particle_temp} is proportional to the adiabatic combustion temperature. The Sherwood and Nusselt numbers can be approximated as $\mathrm{Sh} \approx \mathrm{Nu} \approx 2$, assuming negligible Reynolds number. Since all gas mixtures are tested at the same ambient temperature, Equation~\ref{eq:particle_temp} indicates that the particle temperature depends not only on the adiabatic combustion temperature but is also inversely proportional to the Lewis number. %The temperature difference observed experimentally between the 40\%O$_2$–60\%He and 40\%O$_2$–60\%Ar mixtures -- despite having similar adiabatic combustion temperatures --suggests that the Lewis numbers of these mixtures are not identical. Therefore, further investigation into the mass and thermal diffusivity of these gas mixtures across a range of temperatures is required. 

To compute the thermal diffusivity for each gas mixture ($\alpha(T) = \lambda(T)/[\rho(T) C_\mathrm{P}(T)]$), the temperature-dependent thermal conductivity $\lambda(T)$ is estimated using the empirical Chapman–Enskog relation \cite{kee_chemically_2017}. The density $\rho(T)$ is calculated using the ideal gas law, and the specific heat capacity $C_\mathrm{P}(T)$ is determined from the NASA polynomial expression \cite{mcbride_nasa_2002}. The binary mass diffusivity of oxygen in each inert gas, $D_{\mathrm{O}_2\mathrm{-inert}}(T)$, is also estimated using the Chapman–Enskog relation \cite{kee_chemically_2017}. Using the binary diffusivity calculated, the oxygen diffusivity in each diluent with 40\%O$_2$ gas mixture can then be calculated using Blanc’s law \cite{poling_properties_2007, katz_property_2011}. 

The computed temperature dependencies for both thermal and mass diffusivity are shown in Fig.~\ref{fig:leCANTERA}. The thermal diffusivity of each gas mixture increases with temperature, following a power law, $\alpha \propto T^n$, where $n$ ranges from 1.60 to 1.65. Similarly, the mass diffusivity of oxygen in each gas mixture follows, $D_{\mathrm{O}_2\mathrm{-mix}} \propto T^n$, with $n$ between 1.65 and 1.69, consistent with predictions from the Chapman–Enskog relation that accounts for the temperature dependence of Lennard-Jones potentials governing molecular interactions \cite{kee_chemically_2017}. Because the exponents $n$ are similar for both thermal and mass diffusivity, the Lewis number remains a weak function of temperature, as illustrated in Fig.~\ref{fig:leCANTERA}. The mass diffusivity of oxygen in the 40\%O$_2$–60\%He mixture is notably lower than the binary diffusivity of oxygen in helium. This is due to the substantial reduction in oxygen diffusivity when a large proportion (40\%) of the smaller helium atoms are replaced by much larger oxygen molecules, which reduces the oxygen diffusivity in the 40\%O$_2$–60\%He gas mixture and subsequently increases the Lewis number ($\mathrm{Le} = \alpha / D_{\mathrm{O}_2\mathrm{-mix}}$) well above unity. On the other hand, the molecular diameters and atomic masses of oxygen, argon, and nitrogen do not differ significantly (compared to helium), so substituting either argon or nitrogen with any oxygen concentration in the mixture does not substantially alter the mass diffusivity of oxygen in the mixture. Additionally, because the thermal diffusivity of oxygen, argon, and nitrogen are also similar, the Lewis numbers of N$_2$–O$_2$ and Ar–O$_2$ mixtures should remain close to unity at any oxygen concentrations. Due to the similar Lewis numbers between the argon- and nitrogen-diluted mixtures, the difference in the measured peak temperature in these mixtures is not significant. By comparison, the significantly higher Lewis number in the helium-diluted mixture (Le = 1.82 at 2800\,K) leads to a noticeably lower droplet temperature according to Equation \ref{eq:particle_temp} compared to the argon-diluted case at the same 40\% oxygen concentration. The significance of this temperature difference is also reflected in the particle luminosity shown in the sample frames in Fig.~\ref{fig:tempNsize}, which are captured under identical exposure settings. It is observed that the droplet burns less brightly in the helium-diluted mixture than in the nitrogen- or argon-diluted mixtures, and similar luminosity levels are observed between the nitrogen- and argon-diluted cases.

\subsection{Particle size evolution \label{subsec:size}}

Throughout the combustion process, the size of the burning silicon particle is observed to decrease rapidly over time. The measured peak combustion temperatures, discussed in Section~\ref{subsec:temp}, are well below the boiling point of silicon (3505\,K at 1 atm). According to Glassman’s criterion~\cite{glassman_chapter_2015}, silicon particles are expected to burn heterogeneously as liquid droplets. These liquid droplets are spherical in shape, as shown in the sample frames in Fig.~\ref{fig:tempNsize}, due to the surface tension of molten silicon particles once the particle reaches its melting point (1687\,K) following laser ignition.

Since the silicon particles do not reach their boiling point during combustion, the observed size reduction is primarily attributed to the formation of gaseous SiO (with boiling point of 2150\,K) as an intermediate product species because the combustion temperatures predicted in Section \ref{subsec:temp} favour the formation of SiO compared to SiO\textsubscript{2}. Due to the decreasing particle size over time, a gradual temperature decline after the peak temperature is observed in all experimental cases because the heat release rate from the burning droplet falls more rapidly than the heat loss rate as the particle size decreases \cite{soo_combustion_2018}.

% This is explained by the evolving ratio of heat loss $Q_L$ to heat generation $Q_G$, where $Q_L \propto A_p/r_p = r_p$, $Q_G \propto A_p \propto r_p^2$, and thus $Q_L / Q_G \propto 1/r_p$, as described in Equation \ref{eq:particle_temp}. A sharp temperature drop is observed near the end of combustion, when the particle has shrunk to a sufficiently small size.

The combustion mechanism of silicon is likely similar to that of carbon, as both elements have four valence electrons. In carbon combustion, Asheruddin et al. \cite{asheruddin_analysis_2022} demonstrated good agreement between experimental data and the classical $d^2$-law for biochar combustion at ambient temperatures below 473\,K and oxygen concentrations ranging from 20\% to 100\%. By analogy, silicon combustion is also expected to follow the $d^2$-law. Indeed, a clear linear relationship is observed between the evolution of the squared particle diameter $d^2$ and time $t$, as shown in Fig.~\ref{fig:tempNsize} for all gas mixtures, with a high coefficient of determination ($R^2$ > 0.99). Notably, the slope of the linear fitted curve also remains consistent for the same gas mixture, regardless of the initial particle size, with a standard error of the mean below 2.5\% within each mixture. This strong and repeatable linear relationship between $d^2$ and $t$ indicates that the combustion of single silicon particles follows the relationship of the classic $d^2$ law of combustion as described in Equation \ref{eq:d2law} for a diffusion-controlled combustion. To the authors’ knowledge, such a robust $d^2$-law correlation (with $R^2$ > 0.99) has not previously been reported for other metal fuels. High-speed videos from the experiments reveal that the luminosity of the burning silicon droplet appears uniform across the entire particle surface, which suggests a negligible degree of asymmetric combustion caused by oxide cap formation. This behaviour, combined with the continuous shrinkage of the particle and its negligible density change due to oxidation, likely makes a linear relationship between $d^2$ and $t$ more apparent than in other previously studied metals, such as aluminum or iron. 

\subsection{Burning rate and combustion time}
The slope of the linear fit from the $d^2$-vs-$t$ curve discussed in Section~\ref{subsec:size} reflects the mass transport in different oxidizing environments, and consequently, the burning rate of the droplet. When comparing cases with varying oxygen concentrations using nitrogen as the diluent, while the peak temperature increases by only 337\,K (12\% rise) from air to 100\%O\textsubscript{2}, the slope of the fitted experimental $d^2$-vs-$t$ curves increases substantially, from 127 to 657 \textmu m\textsuperscript{2}/ms (about 5.2 times higher), as shown in Fig.~\ref{fig:tempNsize}. This significant increasing trend is expected, as the oxygen concentration in the bulk gas serves as the driving force for mass transport in diffusion-controlled combustion.

Comparing the gas mixtures of 40\%O$_2$–60\%He and 40\%O$_2$–60\%Ar, a higher burning rate is theoretically expected in the helium-diluted case, since the burning rate is proportional to the oxygen diffusivity, which is higher in helium than in argon gas. However, the experimental results in Fig.~\ref{fig:tempNsize}(d,e) show the opposite trend: the argon-diluted mixture exhibits a higher burning rate. This indicates that the higher combustion temperature in the argon-diluted mixture, as discussed in Section~\ref{subsec:temp}, enhances the droplet regression rate. Although the combustion temperature remains below silicon’s boiling point, the combustion process is likely not purely heterogeneous because of liquid silicon evaporation. According to the Clausius–Clapeyron relation and previous experimental data~\cite{hultgren_selected_1973, mondal_vapor_2022}, silicon vapour pressure increases exponentially with temperature. The presence of silicon vapour has also been confirmed by UV emission spectra in earlier silicon dust flame experiments~\cite{heng_silicon_2025}. Evaporation of silicon can form a lifted flame with an increased reactive surface area. Moreover, vapour-phase combustion shortens the oxygen transport distance, as oxygen can react with silicon vapour before reaching the particle surface, thereby accelerating the overall reaction rate. In addition to silicon vapour, elevated combustion temperatures can also promote the evaporation of SiO\textsubscript{2}, which may be present on the particle surface and acts as a barrier to heterogeneous surface reaction. This SiO\textsubscript{2} film may originate from residual oxide layer formed during the early oxidation stage and from reaction products deposited on the particle surface during combustion, but SiO\textsubscript{2} is estimated to constitute less than 10\% of the products predicted by thermodynamic equilibrium. The evaporation flux of Si and SiO\textsubscript{2} into vacuum, $n_\mathrm{i}''$, depends on their vapour pressure and is given by:

\begin{equation}
n_\mathrm{i}'' = \frac{\alpha_\mathrm{e} P_\mathrm{i}}{\sqrt{2 \pi M_\mathrm{i} R T}}
\label{eq:evap_rate}
\end{equation}

\begin{figure}[h]
\centering
\includegraphics[scale=0.45]{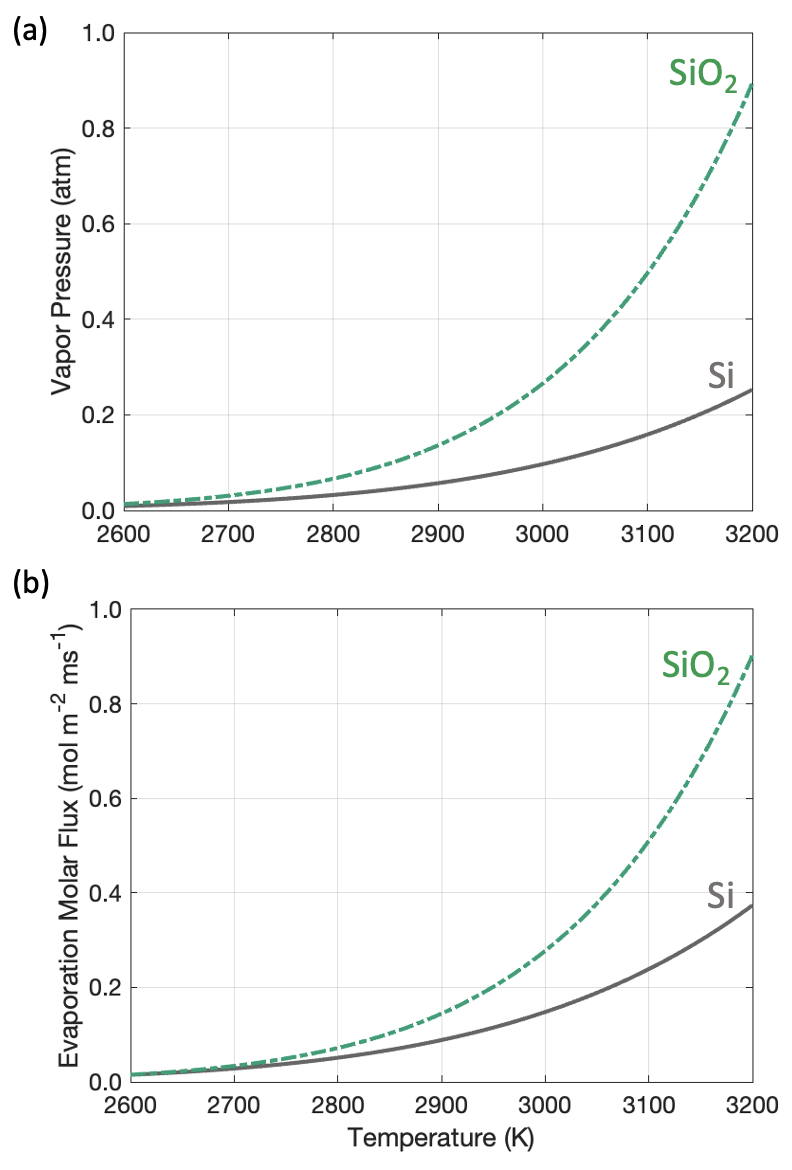}
\caption{(a) Vapour pressure predicted by Clausius–Clapeyron relation at different temperatures and (b) evaporation rate of Si and SiO$_2$ calculated by Equation.~\ref{eq:evap_rate} at different temperatures. 
\label{fig:evap_rate}}
\end{figure}

\noindent Here, $P_\mathrm{i}$ is the vapour pressure of species $i$, $M_\mathrm{i}$ is its molar mass, $R$ is the universal gas constant, $T$ is the droplet temperature, and $\alpha_\mathrm{e}$ is the evaporation coefficient \cite{langmuir_evaporation_1916}. Figure~\ref{fig:evap_rate} shows the estimated vapour pressures of Si and SiO\textsubscript{2} based on the Clausius–Clapeyron relation and the corresponding evaporation rates calculated using Equation~\ref{eq:evap_rate}, assuming $\alpha_\mathrm{e}$ = 1, at various temperatures.

\begin{figure}[h]
\centering
\includegraphics[scale=0.43]{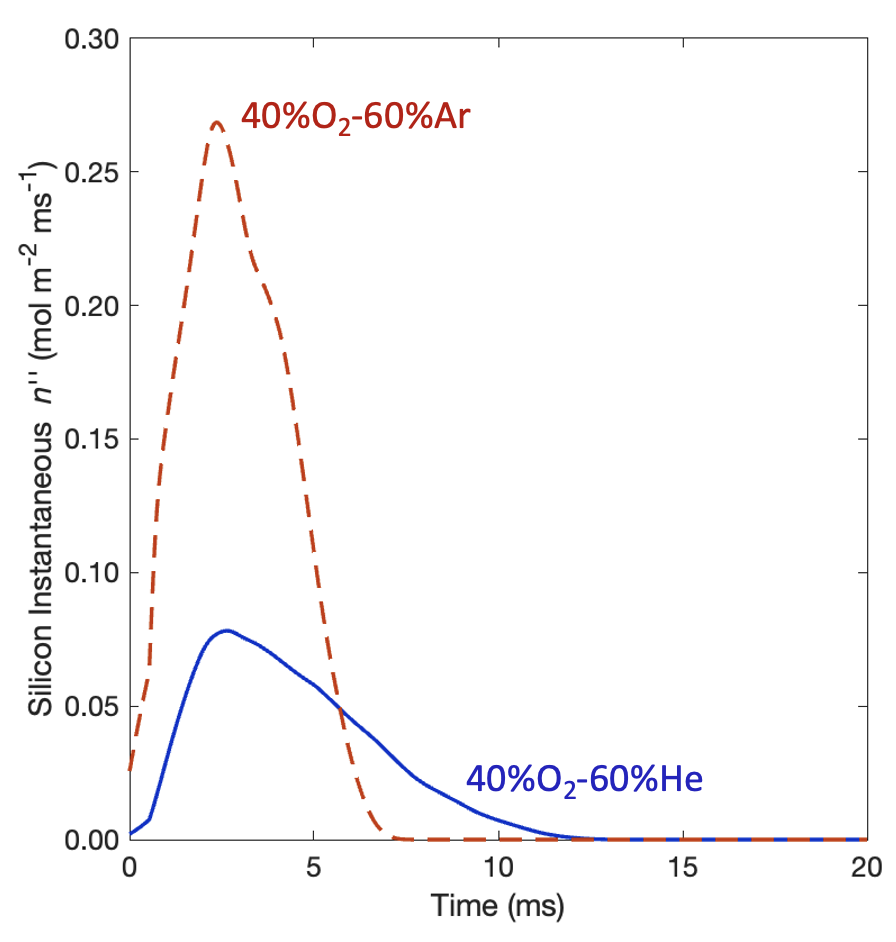}
\caption{Comparison of the instantaneous silicon evaporation flux, $n''$, during droplet combustion in 40\%O$_2$–60\%He and 40\%O$_2$–60\%Ar gas mixtures. The flux profiles are derived from the corresponding droplet temperature histories shown in Fig.~\ref{fig:tempNsize}(d)(e).
\label{fig:evap_rate_compare}}
\end{figure}

The measured peak combustion temperature in the argon-diluted mixture is 249\,K higher than in the helium-diluted mixture. From Fig.~\ref{fig:evap_rate}, this difference corresponds to about a 3.5-fold increase in the silicon evaporation flux at the respective peak temperatures. Figure~\ref{fig:evap_rate_compare} compares the evaporation flux of silicon between the two mixtures as a function of time, based on the temperature profiles in Fig.~\ref{fig:tempNsize}(d,e) and assuming the maximum possible evaporation ($\alpha_\mathrm{e}$ = 1). By integrating the instantaneous flux with the droplet surface area (estimated from the particle diameter) over time, the amount of evaporated silicon throughout the combustion period can be estimated. For an initial particle size of 40-60\,µm, the total evaporated silicon in the argon-diluted mixture is approximately twice that in the helium-diluted mixture, thereby likely contributing to its higher observed burning rate. A similar enhancement of burning rate by liquid-phase evaporation has also been reported by Palecka et al.~\cite{palecka_new_2019}. In their study, it was found that the burning time of iron particles decreased by a factor of three when oxygen concentration increased from 20\% to 40\%, compared to only a twofold reduction predicted by diffusion-limited theory. This was attributed to enhanced gas-phase combustion of iron vapour at higher combustion temperatures with increased oxygen concentration. In Dreizin’s study of aluminum particle combustion~\cite{dreizin_mechanism_1999}, noticeably longer burning times were also observed in He–O$_2$ mixtures than predicted.

\begin{figure}[h]
\centering
\includegraphics[scale=0.45]{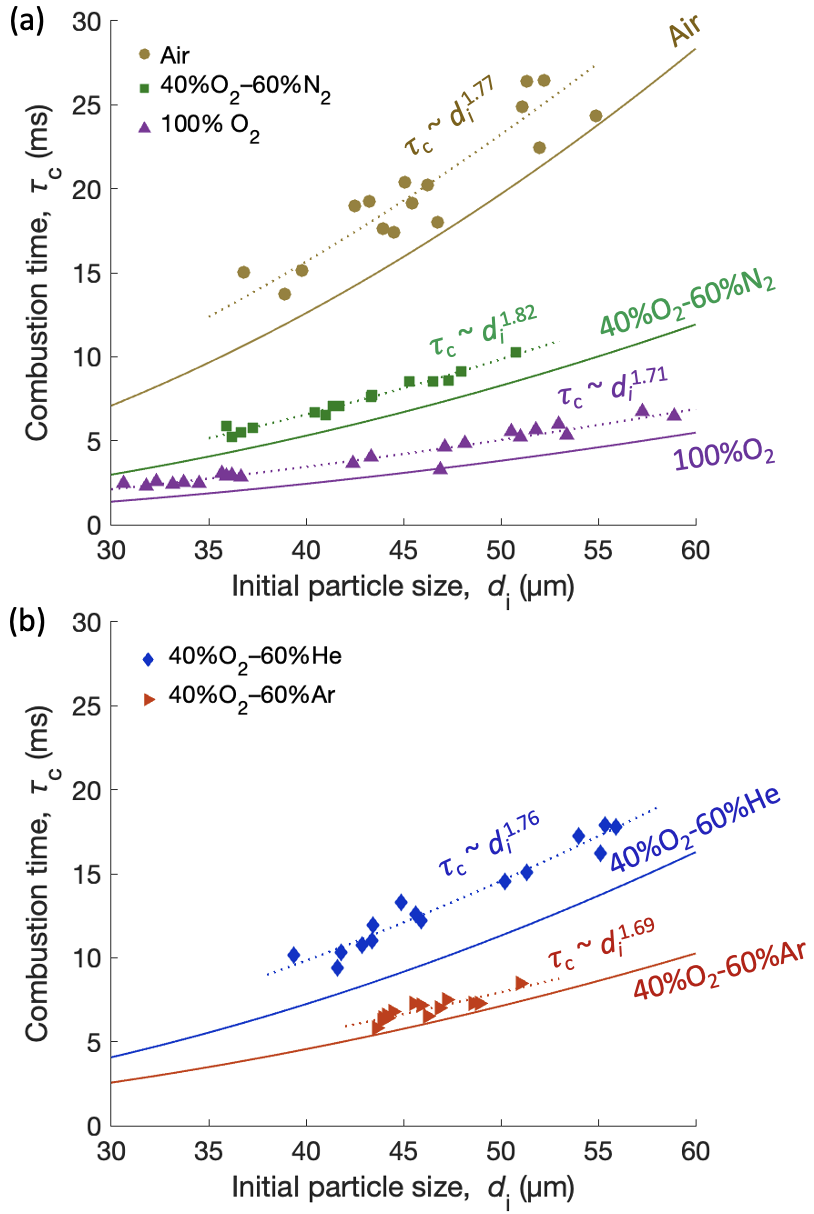}
\caption{Combustion time of silicon droplets, $\tau_\mathrm{c}$, versus initial particle size, $d_\mathrm{i}$, for (a) air, 40\%O$_2$–60\%N$_2$ mixture, and pure oxygen, and (b) 40\%O$_2$ diluted in helium and argon. Scattered data represent experimentally measured values, with uncertainties of $\pm$2.7\,$\mu$m for particle diameter and $\pm$1.3\,ms for combustion time. The dotted lines represent power-law fits of the form $\tau_c \propto d_\mathrm{i}^n$, where the exponent $n$ is obtained by fitting the measured combustion time (scattered data). The solid lines correspond to the classical relationship $\tau_c = d_\mathrm{i}^2 / K$, where $K$ is the measured burning rate obtained from the experiment. \label{fig:tcvsdi}}
\end{figure}

By observing the particle size evolution over time, the complete combustion time of the droplet in each gas mixture can be determined. In this study, the start of combustion is taken at the image frame that shows the particle beginning to glow and turning into a spherical droplet shape. The end of combustion is defined as the frame when the particle size decreases to 10\% of its initial value, beyond which accurate size detection becomes challenging due to limitations in camera resolution. The difference between these two times is defined as the experimental combustion time in the present study. The combustion time, $\tau_\mathrm{c}$, as a function of the initial particle diameter, $d_\mathrm{i}$, for all gas mixtures is shown in Fig.~\ref{fig:tcvsdi}. These measured combustion times are compared to a smooth curve $\tau_\mathrm{c} = d_\mathrm{i}^2 / K$, where the experimentally determined burning rates $K$ from Fig.~\ref{fig:tempNsize} are used. As illustrated in Fig.~\ref{fig:tcvsdi}, the experimentally measured combustion times (scattered points) are slightly longer than the trend of the smooth curve fitted with the same experimental $K$ values. This deviation arises because the droplet does not maintain quasi-steady combustion throughout its entire lifetime. The detected particle diameter near the peak temperature does not differ significantly from the initial particle diameter, indicating that, during the initial 1–5\,ms after the particle begins to melt, a linear $d^2$-versus-$t$ relationship is not yet established. During this transient phase, the rate of change in $d^2$ with time is smaller with a lower burning rate compared to the quasi-steady phase. This observation is consistent with previous droplet combustion experiments involving hydrocarbon fuels~\cite{nuruzzaman_combustion_1971, faeth_fuel_1971}. In the present experiment, particle tracking during the transient phase is, however, particularly difficult because the particle’s low luminosity is close to that of the backlight. Moreover, rapid changes in luminosity from frame to frame make accurate particle size estimation challenging when using a fixed intensity threshold for particle detection.

As shown in Fig.~\ref{fig:tcvsdi}, the fitted trend for the experimental combustion time follows a relationship of $\tau_\mathrm{c} \propto d^n$, with the exponent $n$ ranging from 1.69 to 1.82. The values of $n$ are noticeably lower than 2 despite the strong linear $d^2$-vs-$t$ relationships ($R^2$ > 0.99) observed for each individual run as shown in Fig.~\ref{fig:tempNsize}, and the measured burning time (scattered points) in Fig.~\ref{fig:tcvsdi} also appears to follow the trend of the solid line $\tau_c \propto d_\mathrm{i}^2$. The data scattering observed reflects measurement uncertainties in both the combustion time and initial particle size. Additionally, it should be noted that the particles used in this study come from a single powder batch, so the particle size range is limited to about 30\,\textmu m based on the size distribution shown in Fig.~\ref{fig:PSDnSEM}(a). As such, this limited size range is insufficient to reliably derive the power exponent $n$. Therefore, we suggest that the observed deviation from the classical $\tau_\mathrm{c} \propto d_\mathrm{i}^2$ relationship in the present experiment is simply due to the limitation in the experimental approach and unavoidable measurement uncertainties, as is common in all experimental studies. Based on this observation, it is also possible that these straightforward factors are significant contributors to the discrepancies from the $\tau_\mathrm{c} \propto d_\mathrm{i}^2$ law reported in many previous studies on various fuels, and that the deviations in the power exponent may not be primarily caused by differences in combustion physics between the classical $d^2$ law and the experiments.

\subsection{Combustion products}

\begin{figure}[h]
\centering
\includegraphics[scale=0.55]{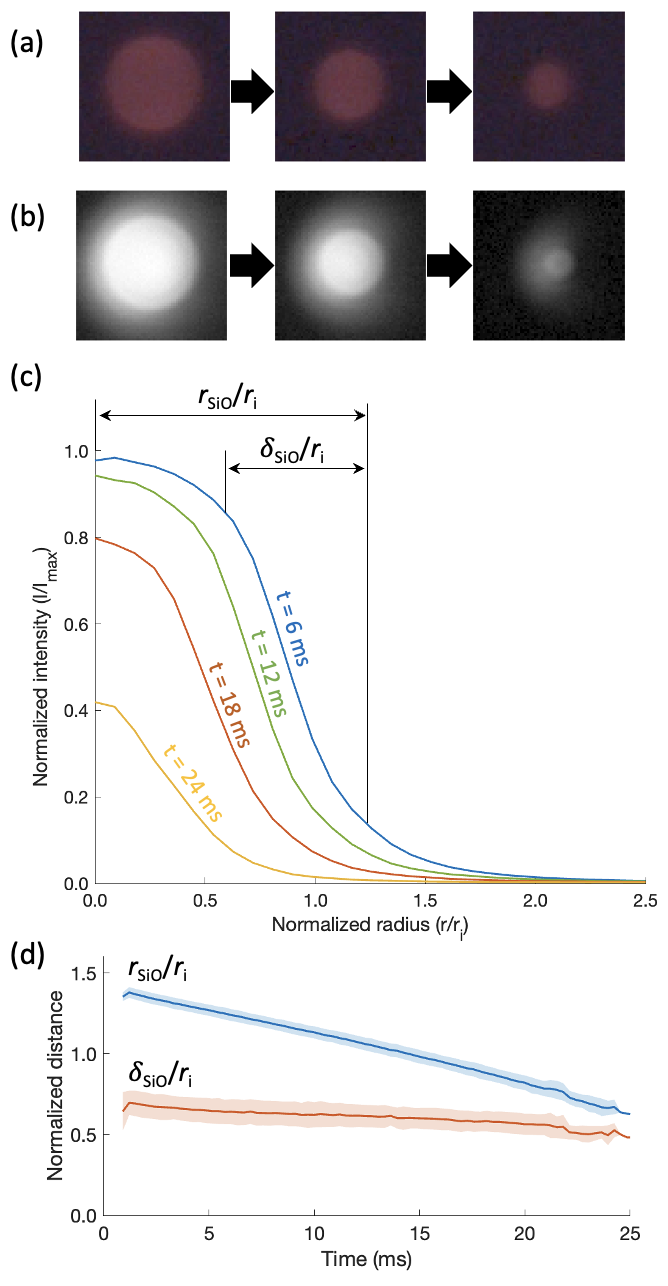}
\caption{Sample frames captured by (a) the colour camera (equipped with a triple bandpass filter at 457, 530, and 628\,nm), and (b) the UV camera (equipped with a bandpass filter ranging from 250–300\,nm). (c) Radial UV intensity profiles of the burning droplet, with intensity normalized by the maximum intensity over the entire combustion period ($I_\mathrm{max}$), and distance normalized by the initial particle radius ($r_i$), such that $r/r_i = 0$ corresponds to the detected droplet centre. Profiles at four representative time steps are shown. (d) Time evolution of the SiO boundary layer edge ($r_{\mathrm{SiO}}$) and its thickness ($\delta_{\mathrm{SiO}}$), both normalized by the initial particle radius ($r_i$). The combustion takes place in ambient air. \label{fig:colour_UV}}
\end{figure}

The rapid decrease in silicon droplet size observed in the experiment is attributed to the formation of gaseous SiO as an intermediate combustion product. Since the emission from excited SiO primarily falls in the UV range \cite{yurchenko_exomol_2021}, the colour camera (sensitive to 400–700 nm) is not suitable to be used to observe this SiO species. Hence, a UV camera coupled with a bandpass filter of 250-300\,nm is employed in the present study to investigate the presence of SiO gaseous species. A comparison of sample frames captured by both the colour and UV cameras is shown in Fig.~\ref{fig:colour_UV}(a) and (b), respectively. The UV images reveal a luminous film surrounding the burning droplet, with noticeable intensity decay away from the particle surface throughout the combustion duration. Figure~\ref{fig:colour_UV}(c) presents the radial UV intensity profile, normalized by the peak intensity throughout the combustion process, at four selected time steps during combustion. An intensity threshold is applied to define the boundary layer edge, $r_{\mathrm{SiO}}$, as the radial position where the normalized intensity falls to 0.15 measured from the particle centre. Accurately detecting the particle surface in the UV images is challenging, as the recorded UV intensity is likely a superposition of SiO emission, Si vapour emission, and blackbody radiation from the liquid droplet. In this study, the particle surface is therefore approximated as the radial position where the normalized intensity reaches 0.85. The boundary layer thickness, $\delta_{\mathrm{SiO}}$, is defined as the distance between the locations corresponding to normalized intensities of 0.85 and 0.15, as illustrated in Fig.~\ref{fig:colour_UV}(c).

% \begin{figure*}[h]
% \centering
% \includegraphics[width=1 \textwidth]{figures/UV_profile.png}
% \caption{Observations captured by the UV camera during silicon droplet combustion in (a) air and (b) pure oxygen. (i) Radial UV intensity profiles of the burning droplet. The intensity is normalized by the global maximum intensity ($I_\mathrm{max}$) over the entire combustion lifetime, and the distance is normalized by the initial particle radius ($r_i$), with $r/r_i = 0$ at the detected centre of the droplet. Profiles at four sample time steps (labelled 1–4) are shown, with their corresponding real-time images. (ii) Time evolution of the SiO boundary layer edge ($r_{\mathrm{SiO}}$) and the boundary layer thickness ($\delta_{\mathrm{SiO}}$), both normalized by the initial particle radius ($r_i$).\label{fig:UV_profile}}
% \end{figure*}

Figure~\ref{fig:colour_UV}(d) shows the time evolution of the boundary layer edge and thickness. The maximum boundary layer edge is observed shortly after ignition, reaching approximately 1.4 times the initial particle radius ($r_{\mathrm{SiO}} \approx 1.4 r_i$). Throughout the combustion process, the boundary layer thickness decreases only gradually and remains around half the initial particle radius ($\delta_{\mathrm{SiO}} \approx 0.5 r_i$). These observations indicate that a significant portion of the UV intensity surrounding the particle originates from gaseous SiO, rather than blackbody radiation emitted by the burning droplet. Previous studies on silicon dust flames have also identified SiO emission in the 260–285\,nm range through spectral analysis \cite{heng_silicon_2025}. While high-speed UV images in the present study are captured only in an ambient air environment, similar UV intensity decay is expected under other oxidizing conditions due to gaseous SiO emission, which is the contributor of the observed reduction in particle size over time. The experimental evidence for gaseous SiO is critical to explain the “limiting combustion temperature” discussed in Section~\ref{subsec:temp}, the formation of SiO$_2$ nanoparticles, and the resulting silicon combustion mode, as hypothesized in \cite{bergthorson_direct_2015} and observed in the previous silicon dust flame experiment \cite{heng_silicon_2025}. The observed UV intensity decay (i.e., SiO decay) is likely associated with a temperature gradient away from the particle surface. As the distance from the droplet increases, the decreasing temperature promotes the condensation of SiO into SiO$_2$, resulting in reduced UV emission intensity. 

% This non-negligible thickness exhibits a distinct UV intensity gradient that is not captured by the colour camera, which primarily records uniform blackbody emission from the liquid droplet. 
% The overall intensity decreases over time due to increasing heat loss relative to heat generation as particle size decreases with time, consistent with the colour camera observations. 
% There is no significant difference in boundary layer thickness between air and pure oxygen environments, likely due to the similar Lewis numbers of both conditions. 

While the formation of condensed SiO$_2$ products is expected, these SiO$_2$ nanoparticles could not be clearly observed with the colour camera for a wide range of exposure times (39–240~\textmu s). The absorption signal from the LED backlight is dominated by noise across the field of view, except at the location of the burning droplet. The extinction coefficient of SiO$_2$ is near zero in the visible and near-infrared (NIR) wavelengths \cite{philipp_silicon_1997}. Previous silicon dust flame experiments have shown that the resulting nanoparticles are smaller than 200\,nm in diameter \cite{heng_silicon_2025}. Based on Mie theory \cite{mie_beitrage_1908}, and given the known refractive index of SiO$_2$, the estimated scattering efficiency for 200\,nm SiO$_2$ particles is below 0.01, and the absorption efficiency is close to zero for the wavelength range of 450-1000\,nm. These results suggest that the nano-sized SiO$_2$ combustion products are effectively transparent to LED white light and therefore not visible in the colour camera. The transparent optical properties of the combustion products, combined with the fact that Si and SiO gaseous emissions occur in the UV range, enable relatively accurate particle tracking in the visible range using the colour camera. This supports the clear linear $d^2$-vs-$t$ relationship observed in the present study (Fig~\ref{fig:tempNsize}). This behaviour contrasts with that of other metal particles, such as aluminum and iron, where the accuracy of droplet size measurement can be compromised by interference from excited gaseous species, visible nano-oxide particles, and oxide caps.

\begin{figure}[h]
\centering
\includegraphics[scale=0.55]{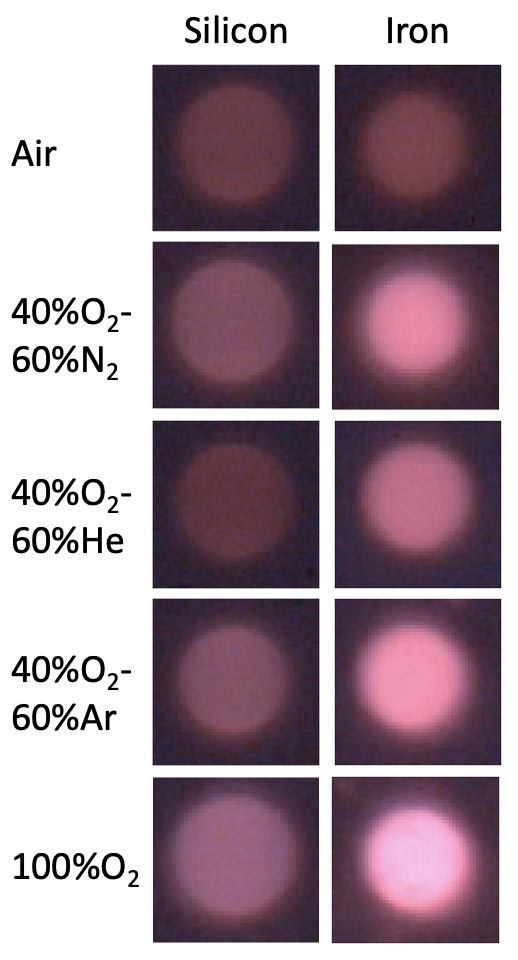}
\caption{Comparison of the peak particle luminosity during combustion of silicon and iron droplets in different gas mixtures. The left column shows silicon particles, while the right column shows iron particles. Each row corresponds to a different oxidizing environment, from top to the bottom: air, 40\%O$_2$-60\%N$_2$, 40\%O$_2$-60\%He, 40\%O$_2$-60\%Ar, and 100\%O$_2$. All images are captured using the same exposure settings. \label{fig:Si_vs_Fe}}
\end{figure}

% Considering the This justifies the grey-body emission assumption applied in the derivation of the temperature profile of the micron-sized burning droplet.
The insignificant absorption of white LED light by the nano-sized SiO$_2$ product implies that the emissivity of the silicon combustion products is likely negligible. As the absorption coefficient of SiO$_2$ nanoparticles is also negligible in the NIR wavelength, the emission signal detected by the photomultiplier is likely dominated by blackbody radiation from the micron-sized burning droplet. Although the present study does not provide a quantitative assessment of the burning droplet emissivity, a qualitative comparison is offered by examining the peak luminosity of silicon and iron droplets under identical oxidizing environments. The peak luminosity for both silicon and iron combustion observed throughout the entire combustion are shown in Fig.~\ref{fig:Si_vs_Fe}. Thermodynamic equilibrium calculations using CEARUN \cite{gordon_computer_1996} indicate that the stoichiometric combustion temperature of silicon is approximately 595\,K higher than that of iron in air, and 250–500\,K higher in environments containing 40\%O$_2$ diluted in helium, argon, or nitrogen. Under 100\%O$_2$, the predicted combustion temperatures for silicon and iron are similar. Despite silicon combustion being expected to produce a higher combustion temperature, the observed luminosity of silicon droplets is significantly lower than that of iron droplets in all oxidizing environments studied under the same camera exposure conditions. This suggests that the emissivity of burning silicon is significantly lower than that of iron, whose emissivity ranges from 0.18 to 0.36, as reported by Yao et al. \cite{yao_spectral_2024} for iron particles in air to pure oxygen diluted in nitrogen, with particle sizes between 44–53 \textmu m. These observations suggest that radiative heat loss from silicon droplet combustion is remarkably lower than that from iron droplets given the same oxidizing environment. For the final condensed combustion products, their transparent nature observed in the present study also indicates a very low emissivity of the nanoparticles, suggesting that radiative heat loss in the post-flame zone of silicon combustion is likely insignificant.

% It is likely that the boundary layer edge corresponds to the distance where almost all gaseous SiO has condensed into SiO$_2$. Thermodynamic analysis indicates that below approximately 2300 K, the equilibrium product of gaseous SiO is less than 0.1\% (by mole), indicating SiO$_2$ formation is strongly favoured over dissociation. Therefore, a sharp temperature gradient exists between the particle surface (2800–3150\,K) and the boundary layer edge (about 2300\,K) over a distance of approximately $0.5r_i$. This sharp gradient limits the available cooling time, preventing the formation of a crystalline structure and favouring the formation of amorphous SiO$_2$, as observed in \cite{heng_silicon_2025}. 

\section{Conclusion}

Single silicon particles are successfully stabilized and ignited in an electrostatic levitator setup, enabling the investigation of their combustion behaviour, including temperature and particle size evolution over time, burning rate, combustion time, and the characteristics of intermediate and final combustion products.

The measured peak temperature of the silicon droplet increases by 337\,K when the oxidizing environment changes from air to pure oxygen. However, the temperature rise with increasing oxygen concentration is relatively insensitive in silicon combustion compared to conventional gaseous fuels (e.g., methane, hydrogen). This low sensitivity is attributed to the fact that the combustion temperature predominantly favours the formation of SiO over SiO$_2$ in the reaction zone. In addition to oxygen concentration, the choice of diluent gas also affects the combustion temperature, even at the same molar oxygen concentration. Although the adiabatic flame temperatures for He–O$_2$ and Ar–O$_2$ mixtures are identical at a given O$_2$ concentration based on thermodynamic analysis, the experiments reveal a noticeably higher flame temperature in the 40\%O$_2$–60\%Ar mixture compared to 40\%O$_2$–60\%He. This difference is attributed to the significantly lower Lewis number in the argon-diluted mixture compared to the helium-diluted one. A higher combustion temperature can significantly enhance the evaporation rates of Si and SiO$_2$, even below their respective boiling points, thereby increasing the burning rate of the silicon droplet.

Across all five oxidizing environments tested, the experiments consistently demonstrate, in each individual run, a strong linear relationship between the square of the particle diameter and real time, with $R^2 > 0.99$. This linearity supports the applicability of the classical $d^2$ law to silicon droplet combustion. However, a deviation from the expected $\tau_\mathrm{c} \propto d_\mathrm{i}^2$ trend was observed, with the fitted exponent $n$ ranging from 1.69 to 1.82. This deviation from the expected $n$ = 2 is likely attributed to measurement uncertainties and the limited particle size range used, rather than differences in combustion physics. %The evaporation of silicon vapour at a higher combustion temperature can affect the burning rates (i.e. proportionality constant), but it is unlikely to significantly affect the power exponent in the classic $d^2$ law as vapour pressure is independent of the particle size. 

% deviate from theoretical estimates. These deviations likely arise from the presence of non-negligible SiO vapour at higher combustion temperatures, which is more pronounced in 40\%O$_2$–60\%Ar, 40\%O$_2$–60\%N$_2$, and pure O$_2$, as well as from the effect of non-unity Lewis numbers in 40\%O$_2$–60\%He. The SiO vapour pressure and the Lewis number of the oxidizing gas are both independent of particle size, so the correction factor for the burning rate is also expected to be independent of particle diameter $d$. This is supported by the clear $d^2 \propto t$ trend observed in each individual run in the experiment for all gas mixtures, despite the deviation in the burning rate from theoretical estimates.
% Developing a more comprehensive model to correct the burning rate for these effects is beyond the scope of this study 
% The Stefan flow of SiO, which resists inward oxygen diffusion, is also expected to be sensitive to temperature.

The formation of gaseous SiO intermediate combustion product is confirmed through UV camera observations, with a SiO film observed that is 40\% larger than the intial particle radius at the beginning of combustion. Furthermore, the condensed combustion products are not detectable via the LED absorption signal, suggesting negligible emissivity and indicating that radiative heat loss from the combustion products is likely minimal in the post-combustion zone. The emissivity of silicon droplets appear to be significantly lower compared to iron droplets. Understanding the combustion behaviour of silicon is essential for advancing its use as an abundant, carbon-free energy carrier. Although this concept has been proposed in prior literature, the present study provides important experimental evidence supporting its technical feasibility to be used as a fuel and deepens the fundamental understanding of its combustion behaviour for the scientific community. 

\section*{CrediT authorship contribution statement}

% \textbf{H. Heng}: Conceptualization, Methodology, Software, Formal analysis, Visualization, Investigation, Validation, Funding acquisition (NSERC-MSFSS), Writing - Original Draft, Writing - Review \& Editing. \textbf{H. Keck}: Methodology, Software. \textbf{C. Chauveau}: Resources, Data Curation. \textbf{S. Goroshin}: Investigation, Writing - Review \& Editing. \textbf{J. Bergthorson}: Project administration, Funding acquisition, Supervision, Writing - Review \& Editing. \textbf{F. Halter}: Investigation, Resources, Funding acquisition, Project administration, Writing - Review \& Editing. 

\textbf{H. Heng}: Conceptualization (lead); Methodology (support); Software (lead); Formal analysis (lead); Visualization (lead); Investigation (lead); Validation (lead); Data Curation (support); Resources (support); Funding acquisition (NSERC-MSFSS); Writing -- Original Draft (lead); Writing -- Review \& Editing (equal). \textbf{H. Keck}: Methodology (lead); Software (support); Resources (support); Writing -- Review \& Editing (equal). \textbf{C. Chauveau}: Data Curation (lead); Resources (lead); Writing -- Review \& Editing (equal) \textbf{S. Goroshin}: Investigation (support); Writing -- Review \& Editing (equal). \textbf{J. Bergthorson}: Project administration (lead -- McGill); Funding acquisition (McGill); Supervision (equal); Writing -- Review \& Editing (equal). \textbf{F. Halter}: Project administration (lead -- CNRS); Supervision (equal); Data Curation (support); Resources (support); Writing -- Review \& Editing (equal). 

\section*{Declaration of competing interest}

The authors declare that they have no known competing financial interests or personal relationships that could have appeared to influence the work reported in this paper.

\section*{Acknowledgments}

This project is funded by the Canadian Space Agency through the Flights and Fieldwork for the Advancement of Science and Technology (FAST) Grant Program (Grant No: 23FAMCGA55). Student work is supported by the Natural Sciences and Engineering Research Council of Canada (NSERC), the Fonds de recherche du Québec – Nature et technologies (FRQNT), and McGill University through the Vadasz Scholars Program. Travel costs associated with the collaboration between McGill University and the Centre national de la recherche scientifique (CNRS) are supported by the NSERC Canada Graduate Scholarships – Michael Smith Foreign Study Supplement (NSERC-MSFSS).

\FloatBarrier

\bibliographystyle{plain}
\bibliography{Zotero_Silicon_Single_Particles}

\begin{thebibliography}{10}
\expandafter\ifx\csname url\endcsname\relax
  \def\url#1{\texttt{#1}}\fi
\expandafter\ifx\csname urlprefix\endcsname\relax\def\urlprefix{URL }\fi
\expandafter\ifx\csname href\endcsname\relax
  \def\href#1#2{#2} \def\path#1{#1}\fi

\bibitem{vaiani_ceramic_2023}
L.~Vaiani, A.~Boccaccio, A.~E. Uva, G.~Palumbo, A.~Piccininni, P.~Guglielmi, S.~Cantore, L.~Santacroce, I.~A. Charitos, A.~Ballini, Ceramic materials for biomedical applications: {An} overview on properties and fabrication processes, J. Funct. Biomater. 14 (2023) 146.

\bibitem{american_coatings_association_use_2020}
{American Coatings Association}, \href{https://www.paint.org/coatingstech-magazine/articles/the-use-of-engineered-silica-to-enhance-coatings/}{The use of engineered silica to enhance coatings}, Raw Materials (Jun. 2020).
\newline\urlprefix\url{https://www.paint.org/coatingstech-magazine/articles/the-use-of-engineered-silica-to-enhance-coatings/}

\bibitem{ai-kazraji_fast_1979}
S.~S. AI-Kazraji, G.~J. Rees, The fast pyrotechnic reaction of silicon and red lead: heats of reaction and rates of burning, Fuel 58 (1979) 139--143.

\bibitem{rugunanan_reactions_1991}
R.~A. Rugunanan, M.~E. Brown, Reactions of powdered silicon with some pyrotechnic oxidants, J. Therm. Anal. 37 (1991) 1193--1211.

\bibitem{koch_special_2007}
E.-C. Koch, D.~Clément, Special {Materials} in {Pyrotechnics}: {VI}. {Silicon} – {An} {Old} {Fuel} with {New} {Perspectives}, Propellants Explos. Pyrotech. 32 (2007) 205--212.

\bibitem{conkling_chemistry_2019}
J.~A. Conkling, C.~Mocella, Chemistry of {Pyrotechnics}: {Basic} {Principles} and {Theory}, 3rd Edition, CRC Press, 2019.

\bibitem{zuo_thermal_2022}
B.~Zuo, S.-Z. Wang, S.~Yang, P.~Liu, Q.-L. Yan, Thermal decomposition and combustion behavior of solid propellant containing {Si}-based composites, Combust. Flame 240 (2022) 111959.

\bibitem{weiser_ignition_2003}
V.~Weiser, E.~Roth, S.~Kelzenberg, A.~Raab, O.~Schulz, Ignition and combustion phenomena of silicon particles in the scope to use it as a green fuel in coal power plants, in: Proceedings of the {European} combustion meeting, 2003.

\bibitem{heng_silicon_2025}
H.~Heng, C.~Mani, N.~Chepel, K.~Mangalvedhe, E.~Antar, S.~Goroshin, J.~Bergthorson, Silicon dust flames: a pathway to using silicon as carbon-free energy carriers, preprint at SSRN (2025).
\newblock \href {https://doi.org/10.2139/ssrn.5182787} {\path{doi:10.2139/ssrn.5182787}}.

\bibitem{turns_introduction_2000}
S.~R. Turns, An {Introduction} to {Combustion}: {Concepts} and {Applications}, 2nd Edition, McGraw-Hill, New York, 2000.

\bibitem{nuruzzaman_combustion_1971}
A.~Nuruzzaman, A.~Hedley, J.~Beér, Combustion of monosized droplet streams in stationary self-supporting flames, Symp. (Int.) Combust. 13 (1971) 787--799.

\bibitem{faeth_fuel_1971}
G.~M. Faeth, R.~S. Lazar, Fuel droplet burning rates in a combustion gas environment, AIAA J. 9 (1971) 2165--2171.

\bibitem{asheruddin_analysis_2022}
N.~M. Asheruddin, A.~M. Shivapuji, S.~Dasappa, Analysis of deviation from classical d2-law for biochar conversion in an oxygen-enriched and temperature-controlled environment, Sci. Rep. 12 (2022) 18391.

\bibitem{marion_studies_1996}
M.~Marion, C.~Chauveau, I.~Gökalp, Studies on the ignition and burning of levitated aluminum particles, Combust. Sci. Technol. 115 (1996) 369--390.

\bibitem{dreizin_mechanism_1999}
E.~L. Dreizin, On the mechanism of asymmetric aluminum particle combustion, Combust. Flame 117 (1999) 841--850.

\bibitem{braconnier_detailed_2019}
A.~Braconnier, C.~Chauveau, F.~Halter, S.~Gallier, Detailed analysis of combustion process of a single aluminum particle in air using an improved experimental approach., Int. J. Energetic Mater. Chem. Propul. 17 (2019) 111--124.

\bibitem{liu_modeling_2021}
J.~Liu, Q.~Chu, D.~Chen, On modeling the combustion of a single micron-sized aluminum particle with the effect of oxide cap, ACS Omega 6 (2021) 34263--34275.

\bibitem{wang_modeling_2021}
J.~Wang, N.~Wang, X.~Zou, W.~Yu, B.~Shi, Modeling of micro aluminum particle combustion in multiple oxidizers, Acta Astronaut. 189 (2021) 119--128.

\bibitem{chen_exploring_2025}
X.~Chen, J.~Liu, Y.~Xu, D.~Zhang, Y.~Tang, B.~Shi, Y.~Feng, Y.~Wu, Q.~Chu, D.~Chen, Exploring the combustion mechanism of single micron-sized aluminum particles with a numerical model, FirePhysChem 5 (2025) 57--67.

\bibitem{ning_burn_2021}
D.~Ning, Y.~Shoshin, J.~Van~Oijen, G.~Finotello, L.~De~Goey, Burn time and combustion regime of laser-ignited single iron particle, Combust. Flame 230 (2021) 111424.

\bibitem{ning_experimental_2024}
D.~Ning, T.~Hazenberg, Y.~Shoshin, J.~Van~Oijen, G.~Finotello, L.~De~Goey, Experimental and theoretical study of single iron particle combustion under low-oxygen dilution conditions, Fuel 357 (2024) 129718.

\bibitem{ning_quantitative_2024}
D.~Ning, A.~Dreizler, A quantitative theory for heterogeneous combustion of nonvolatile metal particles in the diffusion-limited regime, Combust. Flame 269 (2024) 113692.

\bibitem{goroshin_fundamental_2022}
S.~Goroshin, J.~Palečka, J.~M. Bergthorson, Some fundamental aspects of laminar flames in nonvolatile solid fuel suspensions, Prog. Energy Combust. Sci. 91 (2022) 100994.

\bibitem{bar-ziv_electrodynamic_1991}
E.~Bar-Ziv, A.~F. Sarofim, The electrodynamic chamber: {A} tool for studying high temperature kinetics involving liquid and solid particles, Prog. Energy Combust. Sci. 17 (1991) 1--65.

\bibitem{legrand_ignition_1998}
B.~Legrand, E.~Shafirovich, M.~Marion, C.~Chauveau, I.~Gökalp, Ignition and combustion of levitated magnesium particles in carbon dioxide, Symp. (Int.) Combust. 27 (1998) 2413--2419.

\bibitem{keck_new_2024}
H.~Keck, C.~Chauveau, G.~Legros, S.~Gallier, F.~Halter, New experimental method for the simultaneous determination of concentration and size profiles of condensed combustion products around a burning aluminum droplet, Combust. Flame 268 (2024) 113616.

\bibitem{keck_temperature_2025}
H.~Keck, C.~Chauveau, G.~Legros, S.~Gallier, F.~Halter, Temperature field measurement of a burning aluminum droplet, Combust. Flame 277 (2025) 114163.

\bibitem{gordon_computer_1996}
S.~Gordon, B.~McBride, Computer {Program} for {Calculation} of {Complex} {Chemical} {Equilibrium} {Compositions} and {Applications} (1996).

\bibitem{nist_nist_2025}
{NIST}, \href{https://webbook.nist.gov/chemistry}{{NIST} {Chemistry} {WebBook}} (2025).
\newline\urlprefix\url{https://webbook.nist.gov/chemistry}

\bibitem{kee_chemically_2017}
R.~J. Kee, M.~E. Coltrin, P.~Glarborg, H.~Zhu, Chemically {Reacting} {Flow}: {Theory}, {Modeling}, and {Simulation}, 2nd Edition, Wiley, Hoboken, NJ, 2017.

\bibitem{mcbride_nasa_2002}
B.~J. McBride, M.~J. Zehe, {Sanford Gordon}, {NASA} {Glenn} {Coefficients} for {Calculating} {Thermodynamic} {Properties} of {Individual} {Species}, {NASA} {Technical} {Publication} NASA/TP-2002-211556, NASA Glenn Research Center (Sep. 2002).

\bibitem{poling_properties_2007}
B.~E. Poling, J.~M. Prausnitz, J.~P. O'Connell, The {Properties} of {Gases} and {Liquids}, 5th Edition, McGraw-Hill Professional, New York, 2007.

\bibitem{katz_property_2011}
I.~Katz, G.~Caillibotte, A.~R. Martin, P.~Arpentinier, Property value estimation for inhaled therapeutic binary gas mixtures: {He}, {Xe}, {N2O}, and {N2} with {O2}, Med. Gas Res. 1 (2011) 28.

\bibitem{glassman_chapter_2015}
I.~Glassman, R.~A. Yetter, N.~G. Glumac, Chapter 9 - {Combustion} of nonvolatile fuels, in: I.~Glassman, R.~A. Yetter, N.~G. Glumac (Eds.), Combustion, 5th Edition, Academic Press, Boston, 2015, pp. 477--536.

\bibitem{soo_combustion_2018}
M.~Soo, X.~Mi, S.~Goroshin, A.~J. Higgins, J.~M. Bergthorson, Combustion of particles, agglomerates, and suspensions – {A} basic thermophysical analysis, Combust. Flame 192 (2018) 384--400.

\bibitem{hultgren_selected_1973}
R.~Hultgren, P.~D. Desai, D.~T. Hawkins, M.~Gleiser, K.~K. Kelley, Selected {Values} of the {Thermodynamic} {Properties} of {Binary} {Alloys}, American Society for Metals, Metals Park, OH, 1973.

\bibitem{mondal_vapor_2022}
B.~Mondal, T.~Mukherjee, N.~W. Finch, A.~Saha, M.~Z. Gao, T.~A. Palmer, T.~DebRoy, Vapor pressure versus temperature relations of common elements, Materials 16 (2022) 50.

\bibitem{langmuir_evaporation_1916}
I.~Langmuir, The evaporation, condensation and reflection of molecules and the mechanism of adsorption, Phys. Rev. 8 (1916) 149--176.

\bibitem{palecka_new_2019}
J.~Palečka, J.~Sniatowsky, S.~Goroshin, A.~J. Higgins, J.~M. Bergthorson, A new kind of flame: {Observation} of the discrete flame propagation regime in iron particle suspensions in microgravity, Combust. Flame 209 (2019) 180--186.

\bibitem{yurchenko_exomol_2021}
S.~N. Yurchenko, J.~Tennyson, A.-M. Syme, A.~Y. Adam, V.~H.~J. Clark, B.~Cooper, C.~P. Dobney, S.~T.~E. Donnelly, M.~N. Gorman, A.~E. Lynas-Gray, T.~Meltzer, A.~Owens, Q.~Qu, M.~Semenov, W.~Somogyi, A.~Upadhyay, S.~Wright, J.~C. Zapata Trujillo, {ExoMol} line lists – {XLIV}. {Infrared} and ultraviolet line list for silicon monoxide ({28Si16O}), Mon. Not. R. Astron. Soc. 510 (2021) 903--919.

\bibitem{bergthorson_direct_2015}
J.~Bergthorson, S.~Goroshin, M.~Soo, P.~Julien, J.~Palecka, D.~Frost, D.~Jarvis, Direct combustion of recyclable metal fuels for zero-carbon heat and power, Appl. Energy 160 (2015) 368--382.

\bibitem{philipp_silicon_1997}
H.~Philipp, Silicon {Dioxide} ({SiO2}) ({Glass}), in: Handbook of {Optical} {Constants} of {Solids}, Academic Press, 1997, pp. 749--763.

\bibitem{mie_beitrage_1908}
G.~Mie, Beiträge zur {Optik} trüber {Medien}, speziell kolloidaler {Metallösungen}, Ann. Phys. 330 (1908) 377--445, in German.

\bibitem{yao_spectral_2024}
Y.~Yao, D.~Chang, A.~Panahi, Y.~A. Levendis, Spectral emissivities and temperatures of burning iron as single particles or groups of particles, Fuel 375 (2024) 132537.

\end{thebibliography}

\end{document}